\documentclass[
reprint,
superscriptaddress,
amsmath,amssymb,
aps,
]{revtex4-1}

\usepackage{mathrsfs}
\usepackage{graphicx}
\usepackage{dcolumn}
\usepackage{bm}

\usepackage{comment}

\usepackage[hidelinks]{hyperref}

\usepackage{natbib}
\usepackage{color}

\newcommand{\cH}{{\cal H}}
\renewcommand{\Re}{\text{Re}\,}
\renewcommand{\Im}{\text{Im}\,}
\newcommand{\IPR}{\text{IPR}}

\begin{document}

\title{Low energy excitations of mean-field glasses}
\author{Silvio Franz}
\affiliation{\footnotesize LPTMS, UMR 8626, CNRS, Univ. Paris-Sud,
  Universit\'e Paris-Saclay, 91405 Orsay, France}
\author{Flavio Nicoletti}
\affiliation{Dipartimento di Fisica, Universit\`{a} ``La Sapienza'', P.le A. Moro 5, 00185, Rome, Italy}
\author{Federico Ricci-Tersenghi}
\affiliation{Dipartimento di Fisica, Universit\`{a} ``La
  Sapienza'', P.le A. Moro 5, 00185, Rome, Italy}
\affiliation{INFN, Sezione di Roma1, and CNR--Nanotec, Rome unit, P.le A. Moro 5, 00185, Rome, Italy}

\date{\today}

\begin{abstract}
We study the linear excitations around typical energy minima of a mean-field disordered model with continuous degrees of freedom undergoing a Random First Order Transition (RFOT). 
Contrary to naive expectations, the spectra of linear excitations are ungapped and we find the presence of a pseudogap corresponding to localized excitations with arbitrary low excitation energy. Moving to deeper minima in the landscape, the excitations appear increasingly localized  while their abundance decreases.
Beside typical minima, there also exist rare ultra-stable minima, with an energy gap and no localised excitations.
\end{abstract}

\maketitle

\section{Introduction}

The nature of low energy excitations in glasses have attracted a
lot of attention in the last years. Though glasses behave as solids,
disorder induces low energy excitations -both of linear and
non-linear- of of very different nature from the one of the ordered
solids. Remarkably, low energy excitations of glasses display a high
degree of universality. In addition to usual phonons, in a varity of
model glassy system one finds the presence of ungapped low
energy, quasi-localized excitiations with density of states (DOS)
behaving quartically at low frequences
$\rho_{QLS}(\omega)\sim A_4\omega^4$
\cite{lerner2016statistics,mizuno2017continuum,lerner2017effect,shimada2018anomalous,kapteijns2018universal,angelani2018probing,wang2019low, Wang2019sound, richard2020universality,bonfanti2020universal,ji2019,ji2020thermal,ji2021geometry}.
The $\omega^4$ behavior seems to be very
general, independent of the system, preparation protocol and even of
the space dimension. The coefficient $A_4$ on the other hand depends
on the system and the preparation protocol. It appears that deeper
states in the landscape, corresponding to better optimized glasses,
have less and less the low energy excitations, reflecting in smaller and smaller values of $A_4$, and
correspondingly, the excitations are more and more
localized \cite{ji2020thermal,ji2021geometry}.  
This spectrum of localized modes was first rationalized
through phenomenological
theories \cite{gurevich2003anharmonicity,gurarie2003bosonic}, while new predictions have recently enriched the
picture \cite{bouchbinder2021low, rainone2021mean, folena2021marginal,ji2019,ji2020thermal,ji2021geometry, Arceri2020UnificationAtJamming}. 
In addition to typical ungapped minima, 
found by usual minimization protocols, 
it has been noticed in \cite{Kaptjeins2019fastgenultrast} that 
in some model glasses gapped mimima can be found through the use of  smart minimization protocols that include particle swap \cite{BerthierSwapHS2016, BerthierNinnarelloSwap2017}. In such ultrastable minima the $\omega^4$ spectrum is cut-off at low frequencies and localized excoriations are suppressed.

A theoretical comprehension based on microscopic models is however
desirable.  In such context, spin glasses with continuous degrees of
freedom provide a natural playground, the Hessian matrices turn out to
be random matrices from classical ensembles and their spectral
properties can be simply derived. Emblematic is the case of spherical
disordered models where the Hessian belongs to either the Gaussian Orthogonal Ensemble (GOE) ---for instance, the spherical p-spin
models \cite{Crisanti1992, cavagna1998stationary}--- or Wishart ensembles \cite{franz2015universal} (perceptron model), with a constant shift on the
diagonal that ensure that all eigenvalues are positive. In these cases, 
either the minima are gapped and the minimal
excitations have a positive energy, or there is a square-root pseudo-gap,
the spectrum behaves as $\rho(\lambda)\sim\sqrt{\lambda}$ and the
non-linear (spin-glass) susceptibility, associated to the inverse
second moment of $\lambda$ is divergent. In all cases, eigenvectors
are fully delocalized. In a recent paper \cite{franz2021delocalization} we have shown that if one departs from spherical models the situation can be different. 
In a spin glass model with vectorial spins, we showed that stable minima with a 
finite spin glass 
susceptibility, still have low energy quasi-localized excitations, 
resulting in a pseudo-gap in the spectral density. In this paper we generalize 
the analysis to glassy minima of models with a glass transition of the One Replica Symmetry Broken/Random First Order Transition (1RSB/RFOT) kind \cite{parisi2020theory}. These provide good mean-field models of the glass transition and have a finite complexity (configurational
entropy) of stable glassy minima in a finite interval of low energy.
We consider then a natural generalization of the $p$-spin model to vector spins \cite{taucher1992annealedn, taucher1993quenchedn, panchenko2018free}, 
characterize the complexity of the energy minima, and study the spectral properties of the corresponding hessian matrices. We find find that typical stable minima have quasi-localized low energy excitations and no spectral gap. In addition, there are rare ultrastable 
minima where localized excitations are suppressed and the spectrum is gapped. 

The structure  of the paper is the following: in Section II we define the model and study its minima. In Section III we study the complexity as a function 
of the energy. Then we study the spectral density in section IV and the eigenvector statistics in section V. In section VI we study rare ultra-stable minima, where localized excitations are absent. 
Finally, in the Discussion we draw our conclusions. 

\section{The model}

We consider the following version of a $p$-spin model with vector
spins. We have $N$ $m$-dimensional vector variables ${\bm S}_i$ with
$i=1,...,N$ such that $|{\bm S}_i|^2=\sum_{\alpha=1}^m
(S_i^\alpha)^2=1$, interacting through a disordered Hamiltonian 
\begin{eqnarray}
  \label{eq:26}
 {\cal H}[{\bm S}]=-\sum_{p} a_p \sum_{{\bm i},{\bm \alpha}}J_{i_1,...,i_p}^{\alpha_1,...,\alpha_p}S_{i_1}^{\alpha_1}... S_{i_p}^{\alpha_p}
\end{eqnarray}
where the couplings $J_{i_1,...,i_p}^{\alpha_1,...,\alpha_p}$ are
Gaussian variables symmetric over all the indexes but otherwise
independent, with zero mean and variance
$\overline{(J_{i_1,...,i_p}^{\alpha_1,...,\alpha_p})^2}=\frac{p!}{2}
  N^{-(p-1)}$. The model generalizes to $O(m)$ spins the mixed p-spin
  model usually considered for Ising or spherical variables. It
differs from the model considered by Panchenko in \cite{panchenko2018free} 
by the fact that here all the spin components interact with each others,
while in that model only components with the same label interact. This
is a minor difference that does not affect the physics and it is only
for notational simplicity that we choose the present version. As in
the usual mixed p-spin model an alternative formulation of the model,
is provided by defining the Hamiltonian as a Gaussian function with
correlation function 
\begin{eqnarray}
  \label{eq:35}
\overline{ {\cal H}[{\bm S}]{\cal H}[{\bm S'}] } = N f(q({\bm S},{\bm S'}))
\end{eqnarray}
where $q({\bm S},{\bm S'})$ is the overlap 
\begin{eqnarray}
  \label{eq:36}
  q({\bm S},{\bm S'})=\frac{1}{N}\sum_{i=1}^N {\bm S}_i\cdot {\bm S'}_i
\end{eqnarray}
and the function $f$ is
\begin{eqnarray}
  \label{eq:37}
  f(q)=\frac 1 2 \sum_{p} a_p^2 q^p. 
\end{eqnarray}
In this paper we concentrate on the cases $m>2$ and the pure monomial case where a single $a_p$ with $p>2$ is non vanishing.
  
\subsection{Minima of the Hamiltonian}
The equations defining the minima of the model 
state that each spin is aligned with its molecular field: 
\begin{eqnarray}
  \label{eq:38}
  \partial\cH[{\bm S}]/\partial S_i^\alpha+\mu_i S_i^\alpha\equiv \partial\cH_i^\alpha+\mu_i S_i^\alpha=0
\end{eqnarray}
with 
\begin{eqnarray}
  \label{eq:39}
  \mu_i=-{\bm S}_i\cdot \partial\cH_i=|\partial\cH_i|. 
\end{eqnarray}

We will be interested to low temperature linear excitations around
minima of energy $E$. These are ruled by the Hessian matrix.  
The Hessian, which we will implicitly think to be restricted 
to fluctuations orthogonal to each of the ${\bm S}_i$ can be written as
\begin{eqnarray}
  \label{eq:40}
  M_{ij}^{\alpha\beta}=\partial\partial\cH_{ij}^{\alpha\beta}+\mu_i\delta_{ij}^{\alpha\beta}.
\end{eqnarray}
It is well know in these problems \cite{cavagna1998stationary,auffinger2013random} that 
independently of the value $E$ of the energy, the matrix $\partial\partial\cH$ can
be considered as a GOE Wigner-Dyson matrix with random Gaussian i.i.d. elements
with variance $\overline{(\partial\partial\cH_{ij}^{\alpha\beta})^2}=f''(1)/N$. The Hessian $M$ is therefore a random matrix of the
Porter-Rosenzweig (or deformed Wigner-Dyson) ensemble \cite{rosenzweig1960repulsion,brezin1998universal} with elements $\mu_i$
on the diagonal.  Once known the $\mu_i$, the statistical properties of eigenvalues
and eigenvectors can be obtained by the `local resolvent' elements 
$G_{ii}^{\alpha\alpha}(\lambda)=[( M-\lambda+i\epsilon)^{-1}]_{ii}^{\alpha\alpha}$, which verify the
well known equation 
\begin{eqnarray}
  \label{eq:41}
  \sum_\alpha G_{ii}^{\alpha\alpha}(\lambda)=
  (m-1)\frac{1}{\mu_i-\lambda-f''(1) G(\lambda)}
\end{eqnarray}
and $G(\lambda)=\sum_{i,\alpha}G_{ii}^{\alpha\alpha}(\lambda)/N$. Notice that for $\lambda=0$,  $G_{ii}^{\alpha\alpha}(0)$ is just the local susceptibility of the spin ${\bm S}_i$ to an applied field on site $i$. This should be a positive quantity for all $i$ implying that $\mu_i> f''(1) G(0)$ for all $i$ \cite{palmer1979internal,bray1981metastable,bray1982spin,bray1982eigenvalue}. 

In order to study the stability properties of the minima we need
therefore access to the distribution of the molecular fields $\mu_i$. 
Before addressing this task, let us relate the true molecular field moduli
$\mu_i$ to the `cavity fields': that is the molecular fields computed when the
$i$-th variable is removed from the system.

\subsection{A glimpse of the Cavity Method}
At the basis of the application of the ``Cavity Method'' \cite{MPV87} there is the
hypothesis that the solutions to Eq.~(\ref{eq:38}) are continuous upon
removal or addition of a single spin. 
Suppose that a spin configuration ${\bm S}_j$ solves the complete set
of Eq.~(\ref{eq:38}), which includes the coupling with the spin
$i$. Thanks to the fact that couplings are small, we
can use linear response theory 
to relate ${\bm S}_j$ to the corresponding solution ${\bm S}_{j\to i}$
where the spin $i$ is removed. 
We then write 
\begin{eqnarray}
  \label{eq:50}
S_j^\alpha = S^\alpha_{j\to i}
  +\sum_{\beta,\gamma}\chi_{jj}^{\alpha\beta}\partial\partial\cH_{ji}^{\beta\gamma} S_i^{\gamma} 
\end{eqnarray}
which, introducing the cavity field $h_i=|\partial\cH_i({\bm S}_{\to i})|$,
allows us to conclude 
\begin{eqnarray}
  \label{eq:52}
  \mu_i=h_i+f''(1) G_0 \quad\text{with}\quad
 G_0= \frac{1}{N}\sum_{\alpha j}\chi_{jj}^{\alpha\alpha}.
\end{eqnarray}
While Eq.~(\ref{eq:52}) is generally valid for all minima, it does not inform us about the the distribution of the cavity fields and its dependence on the energy level. 
We can obtain this information through the study of the complexity
(configurational entropy) of typical minima with fixed energy $E$.
Notice that Eq.~(\ref{eq:52}) allows to write a self-consistent equation for the resolvent from Eq.~(\ref{eq:41}) that reads
\begin{eqnarray}
  \label{eq:53}
   G(\lambda)=(m-1)\left\langle \frac{1}{h-\lambda-f''(1)[ G(\lambda)-G_0]}\right\rangle
\end{eqnarray}
where the angular average is performed on the (still unknown) distribution of the cavity fields. 
Eq.~(\ref{eq:53}) implies that the susceptibility inside a state is related to the first inverse moment of the field distribution,
\begin{eqnarray}
  \label{eq:54}
  \chi=G_0=G(0)=(m-1)\left\langle \frac{1}{h}\right\rangle,
\end{eqnarray}
while the spin glass susceptibility $\chi_{sg}=\left.\frac{\partial
  G}{\partial \lambda}\right|_{\lambda=0}$ reads
\begin{eqnarray}
  \label{eq:55}
  &&\chi_{sg}=\frac{1}{f''(1)}\frac{1-\Lambda}{\Lambda}\\
&&\Lambda=1-(m-1)f''(1)\left\langle \frac{1}{h^2}\right\rangle
\end{eqnarray}
leading to the stability condition $\Lambda>0$.  It can be shown that $\Lambda$ is the `replicon eigenvalue' appearing in the $T=0$ replica formalism, and whose positivity is necessary for stability.

\section{The Complexity}
According to the theory developed by Monasson in \cite{monasson1995structural}, the complexity of stable states
can be computed through the replica method studying the Replica Symmetric free-energy for non vanishing number of replicas $n$. 
Compared with other existing methods this has the advantage that with the same token one can study both thermodynamics and the properties of the metastable states. 
We need then to consider the average partition function of $n$ replicas at temperature $T=1/\beta$ where all 
the replicas have a mutual overlap $q$:
\begin{equation*}
Z_n=\overline{\int \boldsymbol{dS}\exp\left(-\beta\sum_{a=1}^n \cH[\boldsymbol{S}_a]\right)
\prod_{a,b}\delta(\boldsymbol{S}_a\cdot\boldsymbol{S}_b-N q)}.
\end{equation*}
At the saddle point for $q$, the free-energy as a function of $n$, considered now as a positive real number, is related to the Legendre transform of the complexity of metastable states as a function of the free-energy $g$ by
\begin{eqnarray}
  \label{eq:42}
  {\cal G}(n,T)=\frac 1 N \log Z_n=\Sigma(g,T)-\beta n g
\end{eqnarray}
at the point where $\Sigma'(g)=\beta n$. In order to obtain the
complexity of the energy minima one should consider the limit
$T\to 0$ and $n\to 0$ with $y=\beta n$ fixed:
the result is ${\cal G}_0(y)=\Sigma(E)-y E$. 
A standard calculation that we reproduce in the appendix provides  the expression of the replica symmetric finite $n$ free-energy as follows:
\begin{align}
  \label{eq:43}
{\cal G}(n,T)=&\frac{n \beta^2}{2}\left[  f(1) + (n - 1) (f(q) -  q f'(q)) -  f'(q) \right] \nonumber\\
 &+\log\left[\frac{\int_0^\infty dh\; h^{m-1} e^{-\frac{h^2}{2f'(q)}}Y(\beta h)^n} {\int_0^\infty dh\; h^{m-1} e^{-\frac{h^2}{2f'(q)}}}
 \right]\;,
\\
 &Y(u)= (2\,\pi)^{m/2}\frac{I_{\frac{m-2}{2}}(u)}{u^{\frac{m-2}{2}}}\;,\nonumber
\end{align}
where $I_{\nu}(u)$ is the modified Bessel function of order
$\nu$. The overlap $q$ between the replicas verifies the saddle point equation
\begin{eqnarray}
  \label{eq:45}
  q=\frac{\int_0^\infty dh\; h^{m-1} \exp\left[-\frac{h^2}{2f'(q)}\right] Y(\beta
  h)^{n-2}Y'(\beta h)^2}{\int_0^\infty dh\; h^{m-1} \exp\left[-\frac{h^2}{2f'(q)}\right] Y(\beta h)^n}.\quad
\end{eqnarray}
From the replica free-energy one can also compute the `replicon eigenvalue' $\Lambda$, whose positiveness is a necessary stability condition for the free-energy \eqref{eq:43}. Its expression is rather lengthy and we give it in Appendix \ref{sec:appA}.

Eq.~\eqref{eq:45} has always a trivial $q=0$ solution with vanishing complexity. Depending on the temperature, two $q>0$ solutions can appear. 
The one with a small value of $q$ is always unstable. The one with a larger $q$ can be stable or unstable depending on the sign of $\Lambda$.
From simple thermodynamics, we get the complexity of metastable states at temperature $T$ as a function of the internal free-energy $g$:
\begin{eqnarray}
  \label{eq:44}
g=-\frac 1 \beta \frac{\partial {\cal G}}{\partial n} \qquad
\Sigma=-n^2 \frac{\partial {\cal G}/n }{\partial n}
\end{eqnarray}
The complexity of equilibrium states at temperature $T$ is obtained, as usual, considering the limit $n\to1$ in the previous formulae. Different values of $n$ on the other 
hand, allow to explore different families of metastable states, which have collective vanishing weight at equilibrium. Notice that for fixed $n$ and $T$, the 
present analysis gives us access to the distribution of the cavity field $h$. This distribution can be read 
directly from Eq.~(\ref{eq:43}) and writes:
\begin{eqnarray}
  \label{eq:48}
  P(h)=\frac{ h^{m-1} \exp\left[-\frac{h^2}{2f'(q)}\right] Y[\beta
  h]^n}{\int_0^\infty dh\; h^{m-1} \exp\left[-\frac{h^2}{2f'(q)}\right] Y[\beta
  h]^n}
\end{eqnarray}


The behavior of metastable states is qualitatively similar to the case of the familiar spherical $p$-spin model and follows closely the  RFOT pattern.
The model is paramagnetic at high temperature, Eq.~\eqref{eq:45} has only the $q=0$ solution and the Gibbs measure is concentrated on a single pure state.
Below a dynamical transition transition temperature $T_d$ ergodicity is broken.
In the interval of temperatures $T_K,T_d$ an exponential number of mutually inaccessible metastable states dominate the equilibrium measure: in this situation Eq.~\eqref{eq:45} admits a  stable solution with $q> 0$.
Below $T_K$ the number of states is sub-exponential, the equilibrium measure concentrates on the lowest free-energy states.  
We notice that the replicon eigenvalue, which is vanishing for the states that dominate at $T_d$, is positive at all temperatures below. 

\begin{figure}[t]
    \includegraphics[width=\columnwidth]{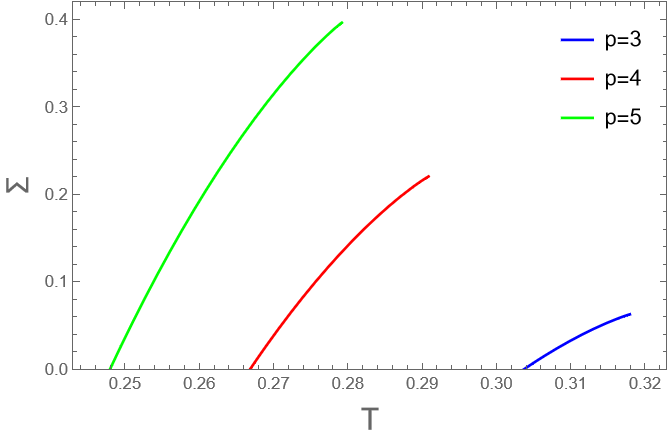}
    \includegraphics[width=\columnwidth]{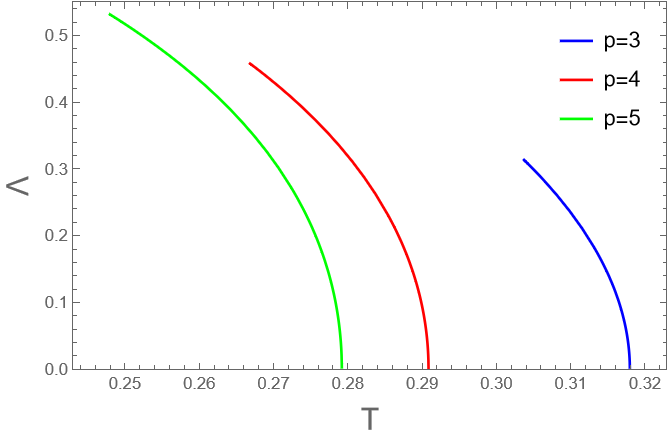}
    \caption{\textbf{Top}: The equilibrium complexity $\Sigma$ for the pure models with $m=4$ and $p=3$ (blue), $p=4$ (red) and $p=5$ (green). The complexity is different from zero in the interval of temperatures $(T_K,T_d)$ and vanishes at $T_K$. The value of the configurational entropy at $T_d$ is $\Sigma_{d}=0.0627787$ ($p=3$), $\Sigma_{d}=0.220444$ ($p=4$), and $\Sigma_{d}=0.396359$ ($p=5$).\newline
    \textbf{Bottom}: The replicon eigenvalue $\Lambda$ for the pure models with $m=4$ and $p=3$ (blue), $p=4$ (red) and $p=5$ (green). The replicon eigenvalue vanishes at $T_d$ as $(T_d-T)^{1/2}$.}
    \label{fig:fig0}
\end{figure}

In Fig.~\ref{fig:fig0} we show the equilibrium complexity and the replicon eigenvalue as a functions of $T$, for $m=4$ and $p=3, 4, 5$. Notice that $\Lambda$ is positive for $T<T_d$ and vanishes at $T_d$ as $\Lambda\sim (T_d-T)^{1/2}$.

\begin{figure}[t]
	\includegraphics[width=\columnwidth]{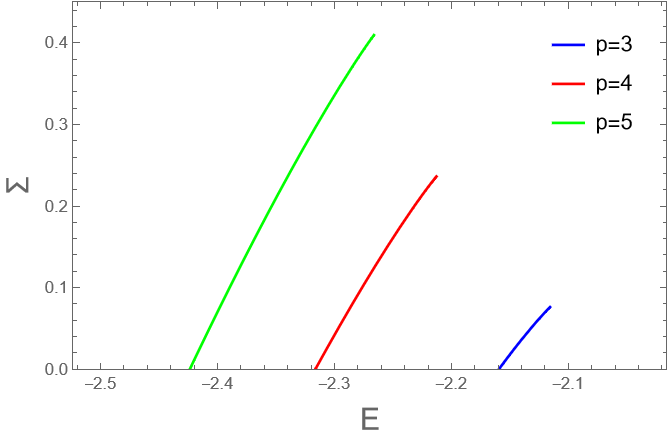}
	\includegraphics[width=\columnwidth]{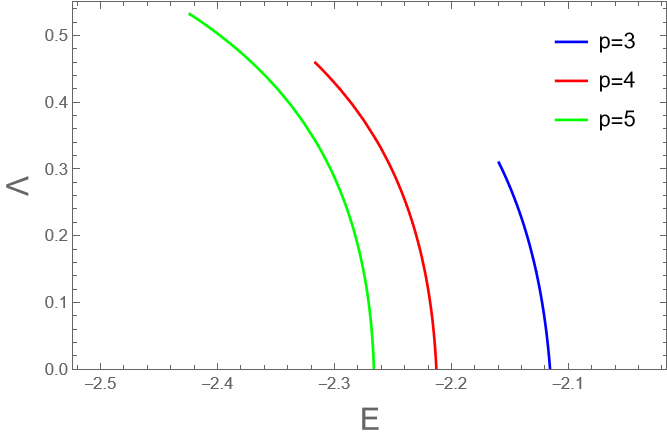}
	\caption{\textbf{Top}: The complexity of the energy minima for the pure models with $m=4$ and $p=3$ (blue), $p=4$ (red) and $p=5$ (green). The maximum complexity is $\Sigma_{max}=0.0760961$ ($p=3$), $\Sigma_{max}=0.236176$ ($p=4$) and $\Sigma_{max}=0.409372$ ($p=5$). The number of stable minima is considerably larger that the number of states at $T_d$.\newline
    \textbf{Bottom}: The replicon eigenvalue in the energy minima for the pure models with $m=4$ and $p=3$ (blue), $p=4$ (red) and $p=5$ (green). Notice that here the replicon eigenvalue vanishes as $E_{mg}-E$, although the slope is very large: we have $|\Lambda'(E_{mg})|\simeq 23, 82, 212$ respectively for $p=3, 4, 5$.
	}
	\label{fig:fig1}
\end{figure}

The number of stable {\it energy} minima can be obtained performing
the limit of $\cal G$ for 
$\beta\to \infty$, $n\to 0$,  keeping the value $y=n \beta$
fixed. In this case, important simplifications occur and, observing that $Y(\beta h)^n\approx e^{y h}$, we get
\begin{multline}
  \label{eq:47}
{\cal G}_0(y)= \frac{1}{2} y^2 \left(f(1)-f'(1)\right) + \\
+\log \left[ \frac{\int_0^{\infty}dh\,h^{m-1}\exp \left(-\frac{h^2}{2 f'(1)}+y h\right)}{\int_0^{\infty}dh\,h^{m-1}\exp \left(-\frac{h^2}{2 f'(1)}\right)}\right]\;,
\end{multline} 
where the last term can be written in terms of confluent hypergeometric functions
\begin{multline}
\frac{\int_0^{\infty}dh\,h^{m-1}e^{-\frac{h^2}{2 f'(1)}+y h}} {\int_0^{\infty}dh\,h^{m-1}e^{-\frac{h^2}{2 f'(1)}}}= {}_1F_1\left(\frac{m}{2};\frac{1}{2};\frac{y^2 f'(1)}{2} \right)+\\
+\frac{\Gamma \left(\frac{m+1}{2}\right)}{\Gamma \left(\frac{m}{2}\right)} y \sqrt{2 f'(1)} {}_1F_1\left(\frac{m+1}{2};\frac{3}{2};\frac{y^2 f'(1)}{2}\right)
\end{multline}
The cavity field distribution in this limit takes the simple form of a reweighed chi distribution:
\begin{eqnarray}
  \label{eq:49}
  P(h)=p_0 h^{m-1} \exp\left[-\frac{h^2}{2f'(1)}+y h\right]
\end{eqnarray}
where $p_0$ is a normalization constant
\begin{eqnarray}
\label{eq:p0}
p_0\,=\,\frac{1}{\int_0^{\infty}dh\,h^{m-1}\,e^{-\frac{h^2}{2 f'(1)}+y h}}\equiv \frac{1}{Z_0}
\end{eqnarray}
The replicon eigenvalue takes exactly the form in Eq.~(\ref{eq:55})
\begin{eqnarray}
  \label{eq:61}
  \Lambda=1-(m-1)f''(1)\left\langle \frac{1}{h^2}\right\rangle
\end{eqnarray}
The study of $\Lambda$ shows that the solution giving the complexity as a function of energy is stable around the ground state energy $E_{gs}$, and only becomes unstable at some higher value $E_{mg}$ of the energy before disappearing at $E_{last}$ \footnote{This is at variance with the Ising case, where all metastable states undergo a Gardner transition at a level specific temperature \cite{Gardner1985, montanari2003nature, GardnerReview2019}}. In order to study the complexity beyond $E_{mg}$ replica symmetry breaking should be included \cite{montanari2003nature,rizzo2013replica}, a task that we will not undertake in this paper. The complexity of the energy minima, within the 1RSB approximation and the corresponding values of the replicon eigenvalue are shown in Fig.~\ref{fig:fig1}.

Comparing Fig.~\ref{fig:fig0} and Fig.~\ref{fig:fig1} we notice that $\Sigma_d < \Sigma_\text{max}$, that is the number of energy minima is much larger than the maximum number of equilibrium states (those dominating the measure at $T_d$).
This feature is at variance to what has been observed in the spherical pure $p$-spin model \cite{castellani2005spin}, where the lack of chaos in temperature preserves the number of states in the whole range of temperatures in the spin glass phase.
Instead it reminds what has been observed in the Ising $p$-spin model \cite{montanari2004cooling} and in the spherical mixed $p$-spin model \cite{Giamp2020}, where the complexity of dominating states may change with the temperature.

\begin{figure}[t]
  \includegraphics[width=\columnwidth]{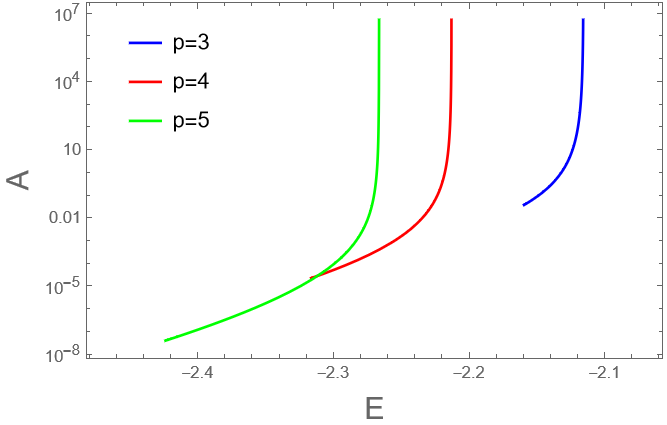}
  \caption{The prefactor $A_4$ of stable glassy minima is smaller for better optimized glasses. The dependence on the energy level $E$ is very strong for high values of $p$: even far from $E_{mg}$ this quantity varies by several order of magnitudes.}
  \label{fig:fig2a}
\end{figure}

\section{The spectral density}
\label{sec:TheSpectralDensity}
We have now all the elements for studying the spectral density of the Hessian matrix in the energy minima from Eq.~(\ref{eq:53}) and Eq.~(\ref{eq:49}).
Let us first make an argument allowing to estimate the
spectrum in the region
\begin{eqnarray}
\Re G(\lambda)-G_0 \ll 1 \;,\quad \Im G(\lambda) \ll 1\;.
\end{eqnarray}
In order to make the argument simpler, let us assume that $m>3$ so that $\left\langle\frac{1}{h^3}\right\rangle <\infty$.
In that region, the leading contribution to the integral in Eq.~(\ref{eq:53}) can be estimated expanding the denominator for small (but non vanishing) values of $\lambda$,
\begin{multline}
  \label{eq:59}
  G(\lambda)\simeq(m-1)\Bigg[ \left\langle\frac1h\right\rangle
  +\left\langle\frac1{h^2}\right\rangle\big[\lambda+f''(1)(G(\lambda)-G_0)\big]+\\
  \left\langle\frac1{h^3}\right\rangle \big[\lambda+f''(1)(G(\lambda)-G_0)\big]^2\Bigg]
\end{multline}
which gives
\begin{eqnarray}
  \label{eq:60}
\rho(\lambda) \propto \Im G(\lambda) \propto &&  
\sqrt{\lambda-\lambda^*}\\
\text{for} \quad \lambda > \lambda^* \equiv
&&\frac{\Lambda^2}{4 (m-1)f''(1)\left\langle\frac 1
 { h^3}\right\rangle}\;.
 \label{eq:lamStar}
\end{eqnarray}
This expression would suggests the existence of a spectral gap $\lambda^*\sim\Lambda^2$ that vanishes only on marginal states where $\Lambda=0$. 
However the expansion in Eq.~(\ref{eq:59}) is not valid for $\lambda\to 0$. 
In fact, any distribution of cavity fields extending its support to $h=0$ is incompatible with a spectral gap, because close to $\lambda=0$ we have $\Re G(\lambda)-G_0=\chi_{SG} \lambda$ and the real part of the denominator in Eq.~(\ref{eq:53}) reads $h-\lambda/\Lambda$. That is, for all the minima but the marginal ones, if we had to admit $\Im G=0$, we would find that the integral in Eq.~(\ref{eq:53}) is divergent. The only possible solution is to have $\rho(\lambda)>0$ for any $\lambda>0$, that is a pseudo-gap for $\lambda<\lambda^*$.

\begin{figure}
	\includegraphics[width=\columnwidth]{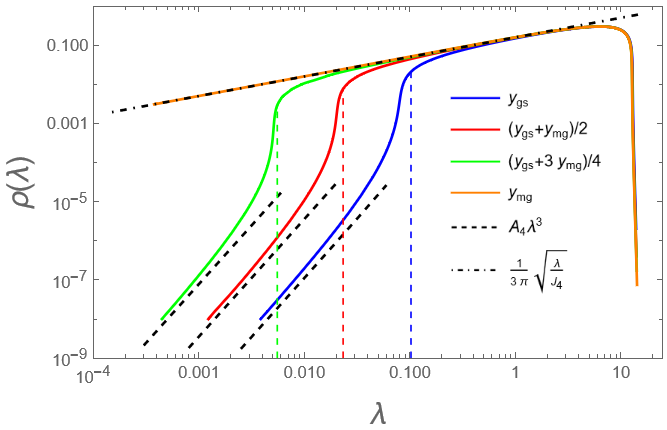}
	\includegraphics[width=\columnwidth]{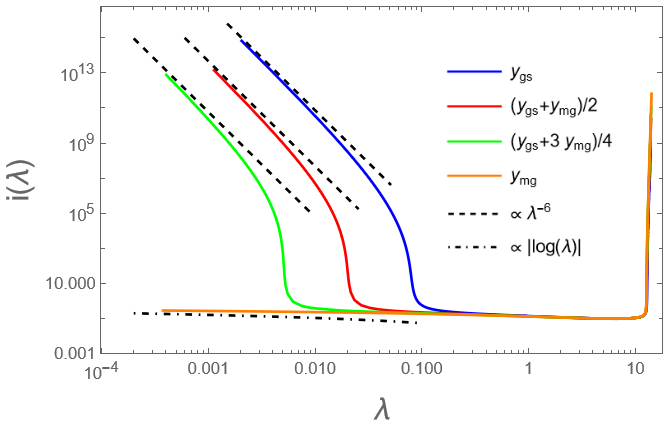}
	\caption{\textbf{Top}: The spectrum of the Hessian in log-log scale for $m=4$ and $p=3$. The curves for $y<y_{mg}$ cross-over from a $\lambda^3$ behavior to a $\sqrt{\lambda}$ behavior at $\lambda_*$ marked by coloured vertical dashed lines. In the bulk of the spectrum, the spectral density does not depend on $y$.\\
    \textbf{Bottom}: The scaled bulk inverse participation ratio $i(\lambda)$ as a function of $\lambda$ for $m=4$ and $p=3$ on a log-log scale. Notice the different behavior between the stable minima and the marginal one. The curve at $y_{mg}$ diverges logarithmically, while the other curves behave as $\lambda^{-6}$ for $\lambda\to 0$.}
	\label{fig:fig2}
\end{figure}

Detailed estimates presented in Ref.~\cite{franz2021delocalization} allow us to conclude that, whenever the field distribution behaves as $P(h) \sim h^{m-1}$ close to the origin (which is the case here), in a stable minimum we have $\Lambda>0$ and a spectral density behaving for small $\lambda$ as 
\begin{equation}
  \label{eq:57}
  \rho(\lambda)=\frac{1}{\Lambda}
  P\left(\frac{\lambda}{\Lambda}\right)\approx A_m\lambda^{m-1}\qquad
  A_m=\frac{p_0}{\Lambda^m}
\end{equation}
This is a pseudo-gap with a power law directly related to the cavity fields `density of states' in the origin and is independent from the energy of the minimum.
The prefactor $A$, conversely, depends on the energy and diverges for $\Lambda\to 0$. Notice that $p_0$ also depends on $\Lambda$ implicitly, since it depends on $y$ which is a function of $\Lambda$. In Fig.~\ref{fig:fig2a} we show the dependence of the prefactor $A_m$ with respect to the energy $E$, in the case of the pure p-spin with $m=4$ and $p=3, 4, 5$. We can see that this term has a strong dependence on the energy, varying by several order of magnitudes in the energy range of the 1RSB landscape. This feature is consistent with what observed for the computer glasses cited in the introduction of this work: the more the minimum is stable and low in energy, the smaller is the prefactor and, consequently, the more localised are the excitations (see discussion below).

As to the case $E=E_{mg}$ or $\Lambda=0$, it was shown in Ref.~\cite{franz2021delocalization} (and we convey the same calculation in Appendix \ref{sec:appB}) that the spectrum behaves as 
\begin{eqnarray}
  \label{eq:58}
  \rho(\lambda)\approx \sqrt{\lambda} \;\;\;\; m>3\\
\rho(\lambda) \approx \sqrt{\frac{ \lambda}{ |\log
  \lambda|}} \;\;\;\; m=3. 
\end{eqnarray}
For finite $\Lambda$, the value $\lambda^*\propto\Lambda^2$, defined in Eq.~\eqref{eq:lamStar}, marks the crossover from the $\lambda^{m-1}$ to the $\sqrt{\lambda}$ behaviors of the spectrum. In Fig.~\ref{fig:fig2} we display the spectrum $\rho(\lambda)$ for $m=4$, $p=3$ and some values of $y$ in the range $[y_{mg},y_{gs}]$ where $y_{mg}=1.42578$ and $y_{gs}=1.94874$. In the plot we check the scaling laws in Eqs.~\eqref{eq:57}, \eqref{eq:58} and show the position of the crossover $\lambda_*$ for each value of $y$. 

\section{The eigenvectors}


The statistics of eigenvectors can be obtained from the study of the
resolvent. It has been shown in \cite{franz2021delocalization} that the eigenvector
components $\psi_i^\alpha$ corresponding to an eigenvalue $\lambda$ in the
bulk of the spectrum are Gaussian variables with a variance given by
\begin{eqnarray}
  \label{eq:33}
  \langle |\psi_i^\alpha|^2 \rangle= \frac{m-1}{N 
  m|h_i+f''(1)(G_0-G(\lambda))-\lambda|^2 }\;.
\end{eqnarray}
where the mean is performed at fixed value of $h_i$ \footnote{For growing size $N$, the cavity fields become uncorrelated to the couplings, and we can therefore treat the off-diagonal elements and the diagonal ones of the Hessian as independent.}.
Notice that the components $\alpha$ are not all independent, as ${\bm \psi}_i$ should be perpendicular to the spin ${\bm S}_i$ in the minimum under consideration. As a result, the Inverse Participation Ratio, $\IPR(\lambda)=\sum_{i\alpha} \langle(v_i^\alpha)^4\rangle$, can be written as 
\begin{eqnarray}
  \label{eq:2}
  &&\IPR(\lambda)=\frac 1 N i(\lambda)= \\
&&  \frac {3(m^2-1)}{N (m+2)}  \int dh \; 
  \frac{P(h)}{|h+f''(1)(G_0-G(\lambda))-\lambda|^4}.\nonumber 
\end{eqnarray}
In the bulk, the IPR is of order $O(N^{-1})$ as it should for a dense matrix. However, close to the edge the eigenvectors are more and more localized. The quantity $i(\lambda)$ grows and diverges at the edges. In particular at the lower edge one can see that 
\begin{eqnarray}
  \label{eq:3}
i(\lambda)\sim \Lambda^3\left(\frac 
  \lambda\Lambda\right)^{-2(m-1)}  
\end{eqnarray}
 for stable minima and 
  \begin{eqnarray}
    \label{eq:27}
    i(\lambda) \propto \left\{
    \begin{array}{ll}
    \sqrt{|\log \lambda|/\lambda} & m=3\\
    {|\log{\lambda}|} & m=4 \\
    \text{const} & m>4
    \end{array}
    \right.
\end{eqnarray}
for the marginal ones. Notice that the minimum eigenvalues
$\lambda_{min}$ are of the order $\lambda_{min}\sim \Lambda
N^{-1/m}$ for stable minima and $\lambda_{min}\sim N^{-2/3}$ for
marginal ones. It is clear that for stable minima Eq.~(\ref{eq:3})
cannot hold till $\lambda\sim \Lambda
N^{-1/m}$, as this would imply an IPR of order $N^{1-2/m}$ which
badly violate the bound $\IPR\le 1$. This suggest that the IPR could
remain finite for the lower eigenvalues, as we will see it is the case
in the next section; we shall then refer to the IPR defined by Eqs.~\eqref{eq:3} and \eqref{eq:27} as \emph{bulk} IPR. For marginally stable minima, { the IPR of the smallest eigenvalue vanishes in the thermodynamic limit, meaning that also the softest modes are delocalised; according to (\ref{eq:27}) the IPR of $\lambda_{min}\sim N^{-2/3}$ goes to zero for $N\to\infty$ as $N^{-2/3}|\ln N|^{1/2}$ for $m=3$, as $N^{-1}\ln(N)$ for $m=4$ and as $N^{-1}$ for $m\geq 5$. In Fig.~\ref{fig:fig2} we show the rescaled bulk IPR, $i(\lambda)$, for $m=4$, $p=3$ and some values of $y$: stable minima have a rapidly diverging $i(\lambda)\sim \lambda^{-2(m-1)}$, whereas at the critical point the divergence is logarithmically slow, in accordance with Eqs.~\eqref{eq:3} and \eqref{eq:27}. Notice that in the case of stable minima the IPRs of lowest eigenvalues should depart from the curves shown at a value $\lambda_*\simeq \Lambda^2$.}

{The necessity of presence of localised excitations in the limit $\lambda\rightarrow 0$ can be understood in a more elegant way, by considering the normalisation condition of eigenvectors given by Eq.~\eqref{eq:33} 
\begin{eqnarray}
 \label{eq:normeig}
  {1=\frac 1 N \sum_i \frac{f''(1)(m-1)}{\left| h_i+f''(1)[G_0-G(\lambda)]-\lambda\right|^2}}
\end{eqnarray}
which is valid for all $\lambda$ in the support of the spectral density. If one assumes that all sites provides a fine contribution to normalisation in the $\lambda\rightarrow 0$ limit, the normalisation condition then would be violated, since for $E<E_{mg}$ the replicon is positive and eq. \eqref{eq:normeig} would imply $1=(m-1)\frac{f''(1)}{N}\sum_i 1/h_i^2$, i.e. $\Lambda=0$. In order to correctly satisfy the normalisation condition at the lower edge, it is necessary to have a condensate component, that yields a finite weight to normalisation in the thermodynamic limit:
\begin{equation}
\label{eq:BE}
    1\,=\,f''(1)(m-1)\left\langle\frac{1}{h^2}\right\rangle+|\Vec{\psi}_C|^2.
\end{equation}
This phenomenon, reminiscent of the Bose-Einstein condensation mechanism, is a very general feature of deformed Wigner matrices \cite{lee2016extremal}}.

\subsection{The spectral edge}
It is interesting to study the statistics of the minimal eigenvalues and their relation
with the low fields. 
This can be done using perturbation
theory \cite{landau1981quantum} around the diagonal matrix, which has the fields $h_i$ as
eigenvalues, which, without loss of generality we will suppose ordered
in increasing order. The low eigenvalues of deep minima are associated 
to sites with small cavity field $h_i$ with $i$ finite for $N\to
\infty$, which for deep minima are such that $h_i\sim N^{-1/m}$ and $h_{i+1}-h_i\sim N^{-1/m}$. 
In fact in correspondence of the lowest
fields $h_i$, one finds multiplets of quasi-degenerate eigenvalues
$\lambda_i^a$, $a=1,...,m-1$ with typical splitting of order
$N^{-1/2}\ll N^{-1/m}$. The eigenvalues can be computed in perturbation theory around the
diagonal matrix ${\rm diag}(\mu_1,...,\mu_N)$, which to the leading order gives \footnote{The same result can be obtained if one considers the condensation condition $|h_i+f''(1)[G(\lambda_i^{a})-G_0]-\lambda_i^a]|=O(N^{-1/2})$ (compare with formula \eqref{eq:normeig}).
}
\begin{eqnarray}
  \label{eq:62}
  \lambda_i^a=h_i+f''(1)G_0+\frac{f''(1)}{N}\sum_{j\ne i} \frac
  {1}{h_i-h_j}\approx \Lambda h_i.
\end{eqnarray}
We obtain for the correspondent eigenvector 
\begin{eqnarray}
  \label{eq:1}
&&  \psi_{k\alpha}^a=\sum_{\beta=1}^{m} \frac{\partial\partial\cH_{ik}^{\alpha\beta} u_{k\beta}^a}{h_k} \;\;\;k\ne i\\
&&\psi_{i\alpha}^a=\sqrt{\Lambda}u_{i\alpha}^a
\end{eqnarray}
where the $m-1$ vectors ${\bm u}_i^a$ are $m$-dimensional
unit norm vectors orthogonal to ${\bm S}_i$ and to each other that at
this level of accuracy in the perturbation theory are left unspecified. Notice that the eigenfunction $\psi$ corresponding to the eigenvalue $\lambda_i^a$ has finite components on the site $i$. The value of the condensate component is in agreement with Eq.~\eqref{eq:BE}.

\section{Ultra-Stable Minima}

\label{sec:UltraStable}

Typical minima are ungapped due to localized excitations associated to sites with small cavity field $h_i$. 
Since the number of minima is exponentially large, one can wonder if rare minima with a gap exist and what is their nature.
In order to search for gapped minima we need to include constraints in the computation of the complexity. Since low energy excitations are related to low 
cavity fields, it is natural to impose a hole in the distribution of the cavity field, $h_i>h_0$ $\forall\; i$ for some $h_0$, which we shall call cavity gap.

The computation of the number of gapped minima is best performed using the Bray-Moore or Kac-Rice formalism \cite{Bray_1981}, computing 
\begin{eqnarray}
  e^{{\cal G}_0(h_0)}&&= \overline{\int_{h_i>h_0} d{\bm S} d {\bm \mu}\;e^{-y{\cal H}}
  \prod_{i,\alpha}\delta\left(
\partial\cH_i^\alpha-\mu_i S_i^\alpha
\right)} \\
&&\overline{\times\left| \det\left( \partial\partial\cH-\text{diag}(\mu)\right)\right|}
  \nonumber
\end{eqnarray}
Since the cavity fields are related to the physical fields $\mu_i=|\partial\cH_i|$ by the equation $\mu_i=f''(1) G_0+h_i$ we impose that $\mu_i>f''(1) G_0 +h_0$. 
The determinant for fixed $\mu_i$ can be computed separately using self-averageness and one can see that 
\begin{eqnarray}
  \overline{\left|\det\left( \partial\partial\cH-\text{diag}(\mu)
  \right)\right|}= e^{\frac{Nf''(1) \chi_{h_0}^2}{2 }}\prod_i [\mu_i-f''(1) \chi_{h_0}]^{m-1}
  \nonumber
\end{eqnarray}
with $\chi_{h_0}$ given by the solution of the saddle point equation \footnote{Notice that, in general, this is not the susceptibility defined by Eq.~\eqref{eq:54}, since in \eqref{eq:chih0} for $h_0>0$ one should integrate from a $\mu_0=h_0+f''(1)\chi>f''(1)\chi$.}.
\begin{eqnarray}
\label{eq:chih0}
  \chi_{h_0}=(m-1) \frac 1 N \overset{N}{\underset{i=1}{\sum}} \,\frac{1}{\mu_i-f''(1)\chi_{h_0}}. 
\end{eqnarray}
The remaining part can be averaged separately and gives
\begin{multline}
\left[
  \frac{1}{\Gamma(m/2)f'(1)^{m/2}}
  \right]^N \exp\Bigg[\frac 1 2 Ny^2f(1)-N\frac { f''(1) u^2}{2}\\ -\sum_i\frac {1}{2f'(1)}[yf'(1)+f''(1) u-\mu_i]^2\Bigg]
\end{multline}
with $u$ given by 
\begin{eqnarray}
\label{eq:u}
  u=\frac{1}{f'(1)N}\overset{N}{\underset{i=1}{\sum}} [\mu_i-yf'(1)-f''(1) u]
\end{eqnarray}
Putting the two terms together, and defining the cavity fields
$h_i=\mu_i-f''(1)\chi$ we obtain
\begin{eqnarray}
&&{{\cal G}_0(y; h_0)= \frac{y^2}{2}[f(1)-f'(1)]-\frac{f''(1)}{2 f'(1)}(\chi-u)^2} \nonumber\\
&& {-f''(1) y (u-\chi)-\frac{f''(1)}{2}(u^2-\chi^2)+\ln I(y; h_0)} \label{eq:g0} \\
&& {I = \frac{\int_{h_0}^{\infty}dh\,h^{m-1}e^{-\frac{h^2}{2 f'(1)}+\frac{h}{f'(1)}[f''(1)(u-\chi)+yf'(1)]}}{\int_0^{\infty}\,dh\,h^{m-1}e^{-\frac{h^2}{2f'(1)}}}} \nonumber
  \nonumber
\end{eqnarray}
Notice that the cavity field probability distribution
\begin{equation}
\label{eq:cavFpdfh0}
    P_{h_0}(h)\,=\,\frac{\theta(h-h_0)}{Z(y; h_0)}h^{m-1}e^{-\frac{h^2}{2 f'(1)}+\left[y+f''(1)\frac{(u-\chi)}{f'(1)}\right]h}
\end{equation}
for $h_0>0$ has a finite cut on the lower edge, that is $P_{h_0}(h_0)>0$, and is re-weighted by the exponential term $y(h_0)\,=\,y+\frac{f''(1)(u-\chi)}{f'(1)}>y$. As a consequence, the gapped minima are therefore more stable than the typical ungapped ones at the same value of $y$, with an energy $E(y;h_0)\,=\,-\partial \mathcal{G}_0(y; h_0)/\partial y$. Different families of ultra-stable minima can be studied by varying $y$ and $h_0$. 


If the lower integration limit is $h_0=0$ it is easy to see by integration by part of \eqref{eq:u} that $\chi=u$, and one gets back \eqref{eq:47} and \eqref{eq:48}. However, this is not the case if $h_0>0$, indeed in such case one finds
\begin{equation}
    u\,=\,\chi_{h_0}+P_{h_0}(h_0).
\end{equation}
In fact, Eq.~(\ref{eq:chih0}), which should be verified substituting the sum by the average over the cavity field distribution, cannot be interpreted as a saddle point condition for the expression in Eq.~(\ref{eq:g0}). The value of $u$ represents linear response of the system to a magnetic perturbation: this quantity, for fixed $y$, is strictly lower than the response $\chi_0$ of the system with $h_0=0$. 
A more detailed discussion of the response in ultra-stable minima can be found in Appendix \ref{sec:appE}.

In the remainder of this section we will discuss the spectral properties and the complexity of ultra-stable minima.  The analytical details behind the formulae we are going to expose
are provided in  Appendices \ref{sec:appD}, \ref{sec:appF}.
 
 As we said, ultra-stable minima have a gapped spectrum, with a lower edge $\lambda_0>0$. It is found for small $\lambda-\lambda_0$ and for small $h_0$
 \begin{eqnarray}
 \label{eq:gap_spec}
   && \rho(\lambda)\propto\sqrt{\lambda-\lambda_0} \\
   && \lambda_0\propto
   \begin{cases}
   \Lambda\,h_0,\qquad\,y>y_{mg} \\
   h_0^{2(m-2)},\qquad\,m>3,\quad y=y_{mg} \\
   h_0^2/|\ln h_0|,\qquad m=3,\quad y=y_{mg}.
   \end{cases} \nonumber
 \end{eqnarray}
{The linear dependence  $\lambda_0=\Lambda h_0$ valid for $y>y_{mg}$ is easily interpreted. It tells that Eq.~\eqref{eq:62} relating small eigenvalues to small fields of typical minima is just cut-off here at the value $h_0$. The localized modes with $\lambda<\lambda_0$ are eliminated without much other effect on the spectrum. 
For $y=y_{mg}$ coherently, the induced spectral gap has a much weaker 
dependence on $h_0$.}

The study of the IPR confirms that in ultrastable minima the most localized are cut-off. 
In presence of a gap $h_0$, the integral appearing in the bulk IPR formula \eqref{eq:2}, remains finite in the limit $\lambda\rightarrow\lambda_0$. By expanding
close to $\lambda=\lambda_0$, it is found at leading order
\begin{equation}
   \begin{cases}
   i(\lambda)\sim h_0^{-2\,(m-1)},\,\,\,y>y_{mg} \\
   i(\lambda)\sim 1/h_0,\,\,\,y=y_{mg},\,\,\,m=3 \\
   i(\lambda)\sim \ln h_0,\,\,\,y=y_{mg},\,\,\,m=4 \\
   i(\lambda)\sim const,\,\,\,y=y_{mg},\,\,\,m\geq 5.
   \end{cases} 
\end{equation}
Details are provided in appendix \ref{sec:appF}.

 In the first panel of Fig.~\ref{fig:spec_gap} we show the spectral density of gapless minima for $m=4$, $p=3$ and $y=(y_{gs}+y_{mg})/2$, comparing it with the spectral density of gapped minima with $h_0=0.15, 0.25, 0.8$: the square root behavior of the spectral edge of ultra-stable minima is confirmed. The spectral density has been computed by solving numerically the following equations
\begin{eqnarray}
\label{eq:spectrum}
& 1\,=(m-1)f''(1)\,\int_{h_0}^{\infty}dh\,\frac{P_{h_0}(h)}{|h+x(\lambda)|^2} \\
& \lambda_0\,=\,f''(1)\chi_{h_0}-x_0-f''(1)(m-1)\left\langle\frac{h+\Re x}{|h+x|^2}\right\rangle_{h_0},\nonumber
\end{eqnarray}
where $x(\lambda)\,=\,f''(1)[\chi_{h_0}-G(\lambda)]-\lambda$. Eqs. \eqref{eq:spectrum} are respectively the imaginary and real part of the equivalent of eq. \eqref{eq:53} when the cavity field PDF is given by Eq.~\eqref{eq:cavFpdfh0}. 

In the second panel of Fig.~\ref{fig:spec_gap} we show, for same $m$ and $p$, the spectral gap as a function of the cavity gap for the values of $y>y_{mg}$ reported in the legend of the plot, comparing the curves with $\Lambda(y)\,h_0$ in each case. The curves were obtained by solving numerically Eq.~\eqref{eq:spectrum} fixing $\lambda=\lambda_0$.
Finally, in the third panel of Fig.~\ref{fig:spec_gap} we show the spectral gap for the case $y=y_{mg}$ and $m=3, 4, 5$, $p=3$, showing the low cavity gap scaling of the $\lambda_0$, which is in good agreement with Eq.~\eqref{eq:gap_spec}

\begin{figure}[t]
     \centering
     \includegraphics[width=0.9\columnwidth]{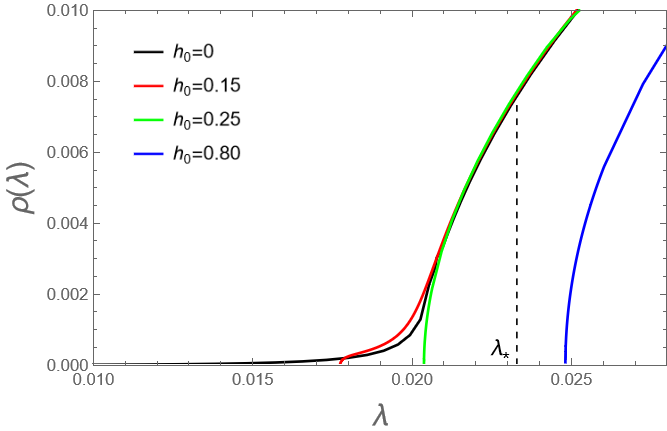}
     \includegraphics[width=0.9\columnwidth]{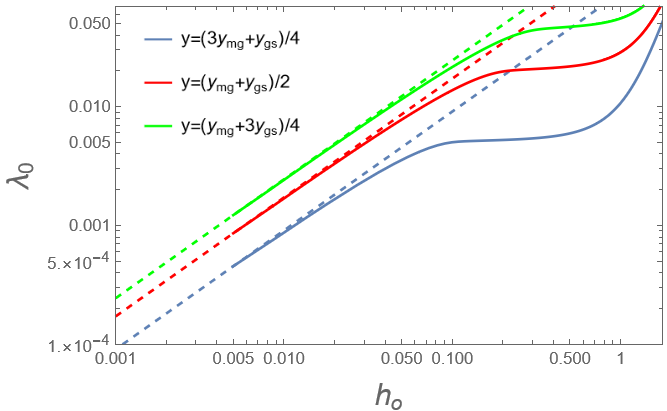}
     \includegraphics[width=0.9\columnwidth]{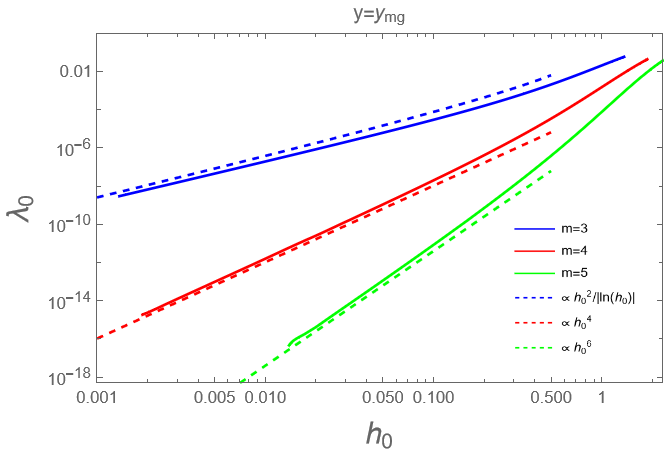}
     \caption{\textbf{Top}: Spectral properties in presence of a cavity gap $h_0$, for the $m=4$ and $p=3$ pure p-spin at $y=(y_{gs}+y_{mg})/2$. The spectral density of gapless minima is compared to that of minima with cavity gaps $h_0=0.15, 0.25, 0.8$. The dashed vertical line marks the position of the crossover $\lambda_*$ in Eq.~\eqref{eq:lamStar}. 
     \newline
     \textbf{Center}: The relation between the spectral gap and the cavity gap for the three values of $y\in[y_{mg}, y_{gs}]$, the dotted lines are $\Lambda(y)\,h_0$. 
     \newline
     \textbf{Bottom}: The spectral gap at the critical point $y=y_{mg}$ for $m=3, 4, 5$: the scaling provided in Appendix \ref{sec:appF} is verified. Marginal minima develop extremely small gaps in a broad range of values of $h_0$.}
     \label{fig:spec_gap}
 \end{figure}

 \begin{figure}[t]
     \centering
     \includegraphics[width=0.85\columnwidth]{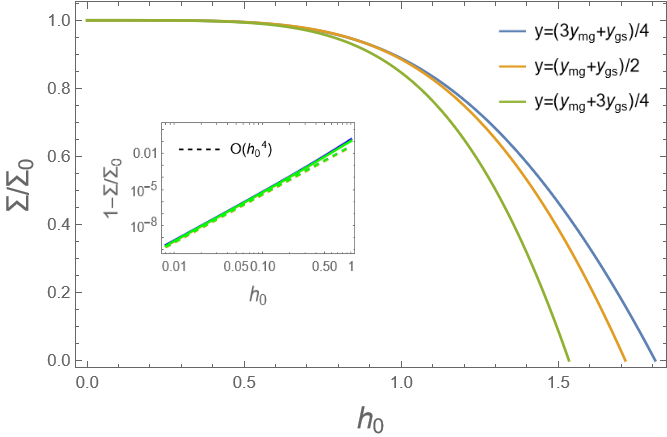}
     \includegraphics[width=0.87\columnwidth]{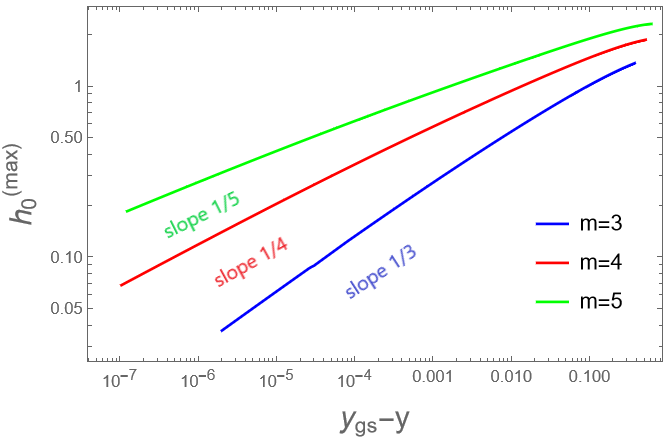}
    \includegraphics[width=0.90\columnwidth]{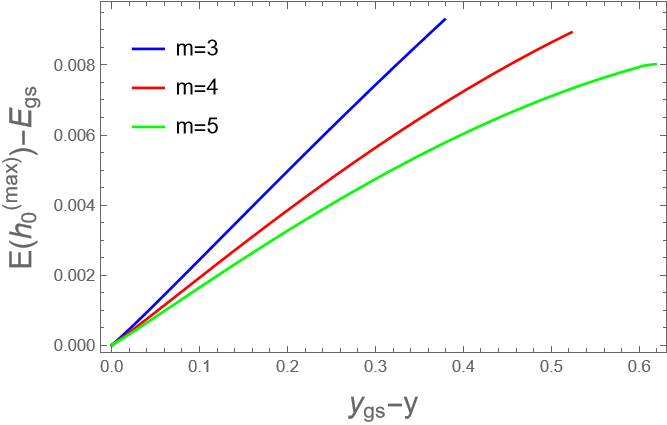}
     \caption{\textbf{Top}: The normalised complexity $\Sigma/\Sigma_0$ for three values of $y\in[y_{mg},y_{gs}]$ and $m=4$: the complexity is a decreasing function of $h_0$, vanishing at a value $h_0^{(max)}(y)$. In the inset, a plot in double log scale of $\Delta\Sigma=1-\Sigma/\Sigma_0$, which shows that for small cavity gap at leading order $\Delta\Sigma=O(h_0^4)$, in agreement with formula \eqref{eq:SigmaGapSmallh0}.
     \newline
     \textbf{Center}: The maximal cavity gap as a function of $y_{gs}-y$ in double log scale, for $m=3, 4, 5$: close to $y=y_{gs}$, this quantity is singular as $(y_{gs}-y)^{1/m}$.
     \newline
     \textbf{Bottom}: The difference between the energy at the maximal cavity gap and the ground state level as a function of $y=y_{gs}$, for $m=3, 4, 5$: there are no ultra-stable configurations down to the ground state.
     }
     \label{fig:complex_gap}
 \end{figure}

 The energy and the complexity of the minima can be computed as usual from ${\cal G}_0(y; h_0)\,=\,-y\,E+\Sigma(E; h_0)$ and $E=-\partial {\cal G}_0(y;h_0)/\partial y$. For any value of $y$, $\Sigma$ is a decreasing function of $h_0$: ultra-stable minima are exponentially small in number with respect to gapless ones. For small cavity gap, the leading behavior is given by
 \begin{equation}
 \label{eq:SigmaGapSmallh0}
     \Sigma=\,\Sigma_0-\left[\frac{1+y\,\langle h\rangle_0}{m\,Z_0}\right]h_0^m+O(h_0^{m+1})
 \end{equation}
 where $\langle\cdot\rangle_0$ is the mean in absence of gap and $Z_0=1/p_0$ (cfr with \eqref{eq:p0}).  In Fig.~\ref{fig:complex_gap} (top) we show the complexity as a function of the cavity gap $h_0$. 
 The complexity is a decreasing function of $h_0$ that vanishes linearly at a value $h_0^{(max)}(y)$. The value of $h_0^{(max)}(y)$ goes to zero as $y$ approaches its value on the ground state of the system. We have in fact $h_0^{(max)}(y)\sim (y_{gs}-y)^{1/m}$: in Fig.~\ref{fig:complex_gap} (center) we check this behavior of the maximal cavity gap for the values of $m=3, 4, 5$.
 As a consistency check, to conclude this section, we show in Fig.~\ref{fig:complex_gap} (bottom) that the energy $E(y; h_0)$ at the maximum cavity gap $h_0^{(max)}(y)$ is always greater than the ground state level $E_{gs}$, for any $y<y_{gs}$: there cannot be ultra-stable minima at the ground state level.

\section{Discussion}

In this paper we have seen that generically, in long range glassy models with continuous variables, stable glassy minima 
posses quasi-localized low energy excitations. In this respect, spherical models, where stable minima are gapped and all excitiations are 
fully extended appear to be the exception rather then the rule. 

We studied the energy minima of a p-spin glass model with $m$-components vector spins. The cases $m=1$ and $m\to \infty$ reduce respectively to the familiar Ising and spherical p-spin models. Similarly to these cases, the model has a 1RSB-RFOT glassy phenomenology, with an exponential multitude  of equilibrium states for temperatures between $T_K$ and $T_d$.

We studied the complexity of the typical minima, which can either be `stable' i.e.\ display a finite spin glass susceptibility, or marginal, with infinite spin glass susceptibility. In this paper we concentrated on the stable minima and the lowest marginal ones, that are described by replica symmetric theories.

Typical minima at each energy level are characterized by a cavity field distribution that extends down to zero. This in turn implies the existence of localized low energy excitations and the absence of a spectral gap. Differently from what observed for models in physical space, the spectrum does not follow a universal $\omega^4$ law. It is still a power law, but the power depends on $m$, the number of components of the vector spins. The prefactor of this power is function of the depth of the minima in the energy landscape, and it is smaller for lower energy. In addition to becoming less numerous, low energy excitations become more and more localized the deeper the minima in the landscape.  Much less numerous than typical minima, also exist rare ultrastable minima where the small fields are absent, localized excitations are suppressed and spectra have a gap. 

In this paper we did not attempt a full characterization of marginal minima. The study of the complexity suggests the existence of marginally stable minima in some intervals of energy above the level $E_{mg}$ that separates stable minima from marginal ones.
These minima are described by replica symmetry breaking and could be the continuation of some high temperature states that undergo a Gardner transition \cite{Gardner1985, montanari2003nature, montanari2004cooling, rizzo2013replica, GardnerReview2019, Scalliet2019marginal} at low temperature. Without much surprise we can expect in these minima a divergent spin glass susceptibility, a square root spectral pseudogap and fully delocalized states.

A natural continuation of this work would be to investigate the spectral properties of low energy excitations of vector spin glass models with finite-connectivity, such as models on random graphs \cite{skantzos2005cavity,coolen2005finitely,marruzzo2015nonlinear,lupo2017approximating,lupo2018comparison,lupo2019random,metz2021mean} or lattice models \cite{baity2015soft,baity2015inherent}. This path would widen our knowledge of the nature of glassy excitations.

\appendix

\section{Computation of the Monasson Free-Energy}

\label{sec:appA}

The computation of ${\cal G}=-\beta n f+\Sigma$ follows standard paths \cite{monasson1995structural}, for completeness we sketch it here the: 
\begin{eqnarray}
Z_n=e^{N{\bf G}}=\overline{\sum_{\bm{S}_a} e^{-\beta \sum_{a=1}^n H({\bm S}_a)}\prod_{a<b}^{1,n}\delta({\bm S}_a\cdot{\bm S}_b-q N)}\nonumber
\end{eqnarray}
where $\underset{\bm S}{\sum}(\cdot)\equiv \overset{N}{\underset{k=1}{\prod}}\,\int d\Vec{S}_k\,\delta(S_k-1)(\cdot)$.
Performing the average and using $\overline{H(\bm S)H(\bm S')}= N f(q_{S,S'})$ one gets
\begin{eqnarray}
&&e^{NG}=\exp\left\{\frac{N\beta^2}{2}[ 
n f(1)+n(n-1)f(q)]
\right\}\zeta(q)\nonumber\\
&&\zeta(q)=\sum_{\bm S_a} \prod_{a<b}^{1,n}\delta({\bm S}_a\cdot{\bm S}_b-q)\nonumber
\end{eqnarray}
The quantity $\zeta$ after one Hubbard-Stratonovich transformation and the integration on spins becomes:
\begin{eqnarray}
  \label{eq:4}
  \zeta&&={\rm St_{\hat{q}}}\;\left\{\exp\left[-N\frac{ n(n-1)}{2} {\hat q} q+\frac{{\hat
  q}}{2}\sum_{a\ne b} {\bm S}^a \cdot  {\bm S}^b\right]\right\}\nonumber\\
&&={\rm St_{\hat{q}}}\left\{\exp\left[-N\frac{ n(n-1)}{2} {\hat q} q -N \frac{n}{2} {\hat q}\right]
 \left[\int D_{\hat q}\Vec{h}\,Y(h)^n\right]^N\right\}.\nonumber\\
&&  Y(h)=(2\,\pi)^{m/2}\frac{I_{\frac{m-2}{2}}(h)}{h^{\frac{m-2}{2}}}\nonumber \\
&& \int D_{\hat{q}}\Vec{h}\,(\cdot)\equiv\int \frac{d\Vec{h}}{(2\pi\hat{q})^{m/2}}\,e^{-\frac{h^2}{2\hat{q}}}\,(\cdot)\nonumber
\end{eqnarray}
Putting everything together and using the saddle point equation
$\hat{q}=\beta^2 f'(q)$ we get (\ref{eq:43}). The physical overlap is found by extremizing $\mathcal{G}$ with respect to $q$ and is given by eq. \eqref{eq:45}: when $T>T_d$, there is only the $q=0$ solution, the system is in a paramagnetic phase with a unique equilibrium state and
\begin{equation}
    \beta g_{para}\,=\,\frac{\beta^2 f(1)}{2}+\log S_m.
\end{equation}
In the range $T_K<T<T_d$, \eqref{eq:45} has a non-trivial solution, corresponding to a non-zero Configurational Entropy: configurations inside the same state have a non-zero overlap, whereas two configurations belonging to two different states have zero overlap. The stability of the non-trivial $q$ is determined by the positiveness of the Replicon Eigenvalue of the Replica Free-Energy Hessian:
\begin{eqnarray}
 \Lambda&&=1-\beta^2f''(q)\Bigl\langle\Bigl\{\frac{m}{(\beta h)^2}\left[\frac{Y'(\beta h)}{Y(\beta h)}\right]^2+ \\
&& \Bigl\{\frac{Y''(\beta h)}{Y(\beta h)}-\left[\frac{Y'(\beta h)}{Y(\beta h)}\right]^2 -\frac{Y'(\beta h)}{(\beta h)\,Y(\beta h)}\Bigr\}^2+\nonumber \\
&& \frac{2 Y'(\beta h)}{(\beta h)\,Y(\beta h)}\times\Bigl\{\frac{Y''(\beta h)}{Y(\beta h)}-\left[\frac{Y'(\beta h)}{Y(\beta h)}\right]^2-\frac{Y'(\beta h)}{(\beta h)\,Y(\beta h)}\Bigr\}\Bigr\rangle\nonumber
\end{eqnarray}
The internal free-energies of TAP states and their Complexity are obtained by eqs.\eqref{eq:45} and they read
\begin{eqnarray}
\label{eq:gS}
& g=-\frac{\beta}{2}[f(1)+(2 n-1) f(q)-(2 n-1)q f'(q) \\
& -f'(q)]-\frac{1}{\beta}\langle\ln Y(\beta h)\rangle_{n}\nonumber \\
& \Sigma=-\frac{n^2\beta^2}{2}[f(q)-q f'(q)]+\ln\zeta-n\langle\ln Y(\beta h)\rangle_{n}
\end{eqnarray}
where $\zeta$ is defined in \eqref{eq:43} and $\langle\cdot\rangle_n$ is an average with respect to \eqref{eq:48}.

Setting $q$ equal to the correct physical value, one can explore different families of metastable states by varying $n$ at fixed $T$ in the range $[T_K, T_d]$, whereas the equilibrium values in the same interval are computed by setting $n=1$. The equilibrium Replicon vanishes at $T_{d}$ as $(T_d-T)^{1/2}$: at higher temperatures, the thermodynamic equilibrium is completely determined by the paramagnetic state $\boldsymbol{m}=0$. The equilibrium Complexity vanishes at $T_K$ as $T-T_K$: for lesser temperature, the Equilibrium Complexity remains zero, meaning that the Gibbs measure is concentrated on the lowest free-energy states.

The $T=0$ limit is performed sending $T$ and $n$ to zero with $y=n/T$ fixed: the result given by eqs. \eqref{eq:47}, \eqref{eq:61} is retrieved by considering the asymptotic expansions of $Y(x)$, $Y'(x)/Y(x)$ and $Y''(x)/Y(x)$:

\begin{eqnarray}
    Y(x) &&\,\overset{x\rightarrow\infty}{\sim}\frac{(2\pi)^{m/2}\,e^x}{x^{m/2-1}}\,\left[\sqrt{\frac{1}{2\pi x}}+O\left(\frac{1}{x}\right)^{3/2}\right] \\
    \frac{Y'(x)}{Y(x)} &&\,\overset{x\rightarrow\infty}{\sim} 1-\frac{m-1}{2\,x}+O\left(\frac{1}{x}\right)^{2} \\
    \frac{Y''(x)}{Y(x)} &&\,\overset{x\rightarrow\infty}{\sim} 1-\frac{m-1}{x}+O\left(\frac{1}{x}\right)^{2}.
\end{eqnarray}



\section{Spectrum of Typical Gapless Minima}
\label{sec:appB}

In this Appendix we will convey the analytical details concerning the spectrum of the energy minima: the analysis is very similar to the one presented in \cite{franz2021delocalization}.\newline
The PDF of the cavity fields moduli at $T=0$, given by \eqref{eq:49}, extends in its support until zero field: as explained in section \ref{sec:TheSpectralDensity}, in this situation the spectrum of the Hessian of $\mathcal{H}$ is necessary gapless. Defining the quantity $x(\lambda)\equiv f''(1)[G(0)-G(\lambda)]-\lambda$, the real and imaginary parts of \eqref{eq:53} satisfy
\begin{eqnarray}
    && \Re G(\lambda)\,=\,\left\langle\frac{h+\Re x(\lambda)}{(h+\Re x(\lambda))^2+\Im x(\lambda)^2}\right\rangle \\
    && 1\,=\,f''(1)\left\langle\frac{1}{(h+\Re x(\lambda))^2+\Im x(\lambda)^2}\right\rangle
\end{eqnarray}
We wish now to consider the $\lambda\rightarrow\,0$ expansion of these equations: to this purpose, we combine them and after some basic rearrangements we get
\begin{eqnarray}
\label{eq:lambdaIJ}
    & \lambda+\Lambda x\,=\,-x^2 J-x |x|^2 I \\
    & J\,=\,f''(1)(m-1)\left\langle\frac{1}{h |h+x|^2}\right\rangle\nonumber \\
    & I\,=\,f''(1)(m-1)\left\langle\frac{1}{h^2 |h+x|^2}\right\rangle\nonumber
\end{eqnarray}
or equivalently
\begin{eqnarray}
\label{eq:IJ}
    &&  I |x|^2=-\Lambda-2 \Re x J\\
    &&\lambda=|x|^2 J \nonumber
\end{eqnarray}
When $\Lambda>0$, the only way to compensate the vanishing of $x$ for $\lambda\rightarrow 0$ in the first of \eqref{eq:IJ} is that $I$ and $J$ are divergent in such limit. For $dG/d\lambda(0)=\chi_{SG}\equiv\frac{1-\Lambda}{\Lambda}$, one has $\Re x(\lambda)\simeq -\frac{\lambda}{\Lambda}$: if $|\Im x(\lambda)|\ll |\Re x(\lambda)|$, one can write $\frac{\Im x}{(h+\Re x)^2+\Im x^2}\approx\pi\delta(h+\Re x)$ and get
\begin{eqnarray}
  \label{eq:16}
&&  J\approx \pi \frac{ \tilde P(|\Re x|)}{|\Re x||\Im x|}\\
&&  I\approx \pi \frac{ \tilde P(|\Re x|)}{|\Re x|^2|\Im x|}\nonumber
\end{eqnarray}
where $\tilde{P}\,=\,f''(1) (m-1) P$ and $P$ is the cavity fields moduli PDF. Plugging these expansions into \eqref{eq:IJ}, we finally get
\begin{eqnarray}
  \label{eq:stab-lowerEdge}
&&  \Lambda=\pi \frac{\tilde  P(|\Re x|)}{|\Im x|}\\
&& J=\Lambda/|\Re x|\nonumber\\
&&\rho(\lambda)=\frac{1}{m-1}|\Im x|/\pi=f''(1) P(\lambda/\Lambda)/\Lambda\sim
   \lambda^{m-1}/\Lambda^m \nonumber
\end{eqnarray}
Eqs. \eqref{eq:stab-lowerEdge} are valid as long as $|\Re x|\ll\Lambda$ and $|\Im x|\ll|\Re x|$, i.e. $\lambda/\Lambda \ll \lambda^{m-1}/\Lambda^m $ or $\lambda\ll \Lambda^{\frac{m-1}{m-2}}$. For $m>3$, a stronger condition is found by considering only $|\Re x|\ll\Lambda$: indeed, we find $\lambda\ll\Lambda^2$; this is equivalent to be at $\lambda\ll\lambda_*$, with $\lambda_*=O(\Lambda^2)$ defined in \eqref{eq:60}.\newline
At the energy level $E=E_{mg}$, we have $\Lambda=0$: eqs. \eqref{eq:IJ} become
\begin{eqnarray}
  \label{eq:IJcrit}
 &&  I |x|^2=-2\,\Re x J\\
&&\lambda=|x|^2 J. \nonumber  
\end{eqnarray}
Integrals $I$ and $J$ now, at variance with $m$, can be finite for $\lambda\rightarrow 0$.
It is easy to see from these last equations that $|\Re x|=\frac{I}{2 J^2}\lambda$ and $|\Im x|\equiv \pi\rho=\sqrt{\lambda/J_0}+O(\lambda)$, with $J_0=f''(1) (m-1) \left\langle\frac{1}{h^3}\right\rangle$: when $I$ and $J$ are finite, it immediately follows $|\Re x|\ll|\Im x|$. Integrals $I$ and $J$ however are finite respectively only for $m>4$ and $m>3$. If $m=4$, $I$ has a logarithmic divergence and $J$ is finite, $|\Re x|\sim\lambda|\ln\lambda|\ll |\Im x|\sim\sqrt{\lambda}$; at $m=3$, if we assume again $|\Re x|\ll|\Im x|$, one finds $I\sim 1/|\Im x|$ and $|\Re x|\sim \sqrt{\lambda}|\ln\lambda|$, $|\Im x|\equiv\pi\rho\sim \sqrt{\lambda/|\ln\lambda|}$. Thus, for any $m\geq 3$ and $\Lambda=0$
\begin{eqnarray}
    & |\Re x|\ll|\Im x| \\
    & \rho\simeq\frac{1}{\pi\,(m-1)}\sqrt{\frac{\lambda}{J_0}},\quad m>3\nonumber\\
    & \rho\simeq\frac{1}{2 \pi\,}\sqrt{\frac{f''(1)\,Z_0\,\lambda}{2|\ln\lambda|}},\quad m=3.\nonumber
\end{eqnarray}

\section{Complexity of Ultra-Stable Minima}
\label{sec:appD}

In this Appendix we show in greater detail all the computations concerning the Complexity of the Ultra-Stable Minima of the energy.
First of all, we set $u-\chi=\Delta$, and rewrite \eqref{eq:g0}
\begin{eqnarray}
\label{eq:G0gapAPP}
& \mathcal{G}_0(y; h_0)= \frac{y^2}{2}[f(1)-f'(1)]-\frac{f''(1)}{2 f'(1)}\Delta^2 \\
& -y f''(1) \Delta-\frac{f''(1)}{2}\Delta(\Delta+2\chi)+\ln \zeta(y; h_0) \nonumber \\
& \zeta_{\Delta} = \frac{\int_{h_0}^{\infty}dh\,h^{m-1}e^{-\frac{h^2}{2 f'(1)}+\frac{h}{f'(1)}[f''(1)\Delta+yf'(1)]}}{{\int_0^{\infty}\,dh\,h^{m-1}e^{-\frac{h^2}{2f'(1)}}}} \nonumber\\
& P_{\Delta}(h)\,=\,\frac{\theta(h-h_0)}{Z(y; \Delta)}h^{m-1}e^{-\frac{h^2}{2 f'(1)}+\left[y f'(1)+f''(1)\Delta\right]\frac{h}{f'(1)}}
  \nonumber
\end{eqnarray}
By combining eqs.\eqref{eq:chih0}, \eqref{eq:u} and approximating the sums with integrals, we find that $\Delta$ satisfies the self-consistent equation
\begin{equation}
    \Delta\,=\,\frac{h_0^{m-1}\,e^{-\frac{h_0^2}{2 f'(1)}+[y+f''(1)\Delta]h_0}}{\int_{h_0}^{\infty}dh\,h^{m-1}\,e^{-\frac{h^2}{2 f'(1)}+[y+f''(1)\Delta]h}}\equiv P_{\Delta}(h_0)
\end{equation}
In particular, for small $h_0$ one has ($Z_0(y)=1/p_0$ defined in eq.\eqref{eq:p0})
\begin{equation}
\label{eq:Deltah0small}
    \Delta\,=\,\frac{h_0^{m-1}}{Z_0(y)}(1+h_0 y)+O(h_0^{m+1}).
\end{equation}
The expression of $\Sigma(y; h_0)$ is obtained by applying the definition $\Sigma(y; h_0)\,=\,y E(y; h_0)+\mathcal{G}_0(y; h_0)$, and the full expression is
\begin{eqnarray}
    & \Sigma(y; h_0)\,=\,\Sigma(y; 0)-\left[\frac{f''(1)^2}{f'(1)}+f''(1)\right] \\
    & \times\left[\frac{1}{2}+\frac{y f'(1) (\langle h \rangle_{\Delta}-h_0)}{f'(1)+(\langle h \rangle_{\Delta}-h_0) f''(1)\Delta}\right]\Delta^2+\nonumber \\
    & -y[\chi+yf''(1)]\frac{f'(1)\,(\langle h \rangle_{\Delta}-h_0)}{f'(1)+(\langle h \rangle_{\Delta}-h_0)f''(1)\Delta}\Delta-\chi\Delta\nonumber \\
    & +\frac{y f'(1) [f''(1)(m-1)-\chi\langle h\rangle_{\Delta}]}{f'(1)+f''(1)(\langle h \rangle_{\Delta}-h_0)\Delta}\Delta\nonumber \\
    & -\frac{y f'(1) [\langle h \rangle_{\Delta}-\langle h \rangle_0]-\langle h \rangle_{0}(\langle h \rangle_{\Delta}-h_0)\Delta}{f'(1)+(\langle h \rangle_{\Delta}-h_0)\Delta}+\ln[\zeta_{\Delta}(y)/\zeta_0(y)]\nonumber
\end{eqnarray}
where $\langle \cdot \rangle_{\Delta}$ is a mean according to $P_{\Delta}$ in \eqref{eq:G0gapAPP}. This nasty expression can be simplified a lot by expanding for low cavity gap: by substituting \eqref{eq:Deltah0small} one gets
\begin{equation}
\label{eq:Sigmah0small}
     \Sigma=\,\Sigma_0-\left[\frac{1+y\,\langle h\rangle_0}{m\,Z_0}\right]h_0^m+O(h_0^{m+1}).
 \end{equation}
For $h_0=O(1)$, $\Sigma$ becomes proportional to $h_0^{(max)}(y)-h_0$, thus vanishing at a certain maximal cavity gap. This last quantity is $O(1)$ far from $y_{gs}$; as this point is approached, the maximal cavity gap is expected to vanish, since  ultra-stable minima cannot be lower in energy than the ground state level. Taking $\Sigma=0$ in \eqref{eq:SigmaGapSmallh0}, we can consider $\Sigma_0$ small and expand it linearly in $y_{gs}-y$, getting
\begin{eqnarray}
    &\left[\frac{1+y\,\langle h\rangle_0}{m\,Z_0}\right](h_0^{max})^m\simeq \frac{d\Sigma_0}{dy}(y_{gs})(y_{gs}-y)\nonumber\\
    & h_0^{(max)}\simeq A\,(y_{gs}-y)^{1/m} \\
    & A\,=\,\left[(m\,Z_0)\frac{\Sigma_0'(y_)}{1+y\langle h \rangle_0}\right]^{1/m}\Bigl |_{y=y_{gs}}
\end{eqnarray}
that is, a singularity approaching $y_{gs}$.

\section{Response Function of Ultra-Stable Minima}
\label{sec:appE}
This appendix is devoted to the computation of the linear response function of the system when perturbed in a ultra-stable configuration at zero temperature: we show that the linear response function in this case is given by the order parameter $u$, which satisfies
\begin{equation*}
    u\,=\,\chi_{h_0}+P_{\Delta}(h_0)
\end{equation*}
Suppose to perturb the system with an external field $\Vec{\epsilon}_i$ on each site: the static linear response function is given by
\begin{eqnarray}
    & \mathcal{R}\,=\,\frac{1}{N}\sum_{i,\alpha}R_{ii}^{\alpha\alpha} \\
    & R_{i j}^{\alpha \beta}\,=\,\frac{\partial \overline{\langle S_i^{\alpha}  \rangle}}{\partial \epsilon_j^{\beta}}\Bigl|_{\epsilon=0}
\end{eqnarray}
where off-diagonal terms of the response matrix are neglected since their disorder average is zero. Here $\langle\cdot\rangle$ is an average according to Kac-Rice-Moore measure:
\begin{eqnarray}
\label{eq:KRM}
    P_{KRM}\,\propto\,e^{-y{\cal H}}
  \prod_{i,\alpha}\delta\left(
{\cal H}_i^{\alpha'}-\mu_i S_i^\alpha
\right)\left| \det\left( H''-\text{diag}(\mu)\right)\right|
  \nonumber
\end{eqnarray}
Then, one has for the response
\begin{eqnarray}
\label{eq:R2}
    & R_{ii}^{\alpha\alpha}\,=\,\langle (S_i^{\alpha})^2 \rangle-\langle S_i^{\alpha} \rangle^2+i\langle S_i^{\alpha} \hat{S}_i^{\alpha} \rangle
    \\ 
    & \rightarrow\mathcal{R}\,=\,\frac{1}{N}\overset{N}{\underset{k=1}{\sum}}\overline{\langle \Vec{S}_k\cdot i\Vec{\hat{S}}_k \rangle}\nonumber
\end{eqnarray}
where $\hat{S}_i^{\alpha}$ are Lagrange multipliers that ensures the $\boldsymbol{S}$ configuration is one of minimum of $\mathcal{H}$ (they are obtained from the Fourier Representation of the delta function in \eqref{eq:KRM}). After performing similar passages to those explained in section \ref{sec:UltraStable}, one finds for the relevant part of the integrals involved in the second eq. of \eqref{eq:R2}
\begin{eqnarray*}
   & \prod_l\int d\Vec{\mu}_l\int d\Vec{\hat{S}}_l\;(\Vec{S}_k\cdot i\Vec{\hat{S}}_k)e^{-\frac{f'(1)}{2}\hat{S}_l^2-i[\mu_l-u-y f'(1)](\Vec{S}_k\cdot i\Vec{\hat{S}}_k)}\propto \\
   & \propto\,-\frac{\int d\boldsymbol{\mu}\frac{\partial e^{-\sum_l\frac{c_l^2}{2 f'(1)}}}{\partial c_k}}{\int d\boldsymbol{\mu}e^{-\sum_l\frac{c_l^2}{2 f'(1)}}}\Bigl|_{c_l\equiv\mu_l-u-y f'(1)}\,=\,\frac{\overline{\mu}-u-y f'(1)}{f'(1)}
\end{eqnarray*}
The remainder of the integrals and factors cancel out with the normalization, and in the end we get
\begin{equation}
    \mathcal{R}\,=\,\frac{1}{N}\sum_{k,\alpha}\,\overline{R_{kk}^{\alpha\alpha}}\,=\,\frac{1}{f'(1)}[\overline{\mu}-u-y f'(1)]\equiv u.
\end{equation}
To conclude this Appendix, we show that $u$ is always smaller than the susceptibility $\chi$ of the typical minimum configurations. From the definition of $\chi_{h_0}$ (eq.\eqref{eq:chih0})
\begin{eqnarray*}
    & \chi_{h_0}\,=\,\chi-\frac{(m-1)\int_0^{h_0}dh\,h^{m-2}e^{-\frac{h^2}{2 f'(1)}+[y f'(1)+f''(1)\Delta]\frac{h}{f'(1)}}}{\int_0^{\infty}dh\,h^{m-1}e^{-\frac{h^2}{2 f'(1)}+[y f'(1)+f''(1)\Delta]\frac{h}{f'(1)}}} \\
    & \equiv \chi-Q(h_0)<\chi
\end{eqnarray*}
one finds
\begin{equation*}
    u\,=\,\chi-[Q(h_0)-P_{\Delta}(h_0)].
\end{equation*}
We notice that $Q(h_0)\,=\,(m-1)\int_0^{h_0}\,dh \tilde{g}(h)$ and $P_{\Delta}(h_0)\,=\,h_0 \tilde{g}(h_0)$, and thus we must determine if $(m-1)\int_0^{h_0}\,dh \tilde{g}(h)-h_0 \tilde{g}(h_0)>0$; this inequality is indeed always verified for $m>2$, since in this circumstance $Q$ is a convex function: we conclude that $u<\chi$. In particular, for small $h_0$ it holds
\begin{equation}
    u\,=\,\chi-\frac{1}{Z_0}[(m-1)(m-2)-1]h_0^{m-1}+O(h_0^m).
\end{equation}

\section{Spectrum of Ultra-Stable Minima}
\label{sec:appF}
When a cavity gap $h_0$ is present, one has a spectral gap $\lambda_0>0$ if the quantity $\Re x(\lambda)\,=\,f''(1)[\chi_{h_0}-G_R(\lambda)]-\lambda$ satisfies $|\Re x(\lambda_0)|<h_0$: in these circumstances, the spectral gap is determined by solving
\begin{eqnarray}
\label{eq:gapSpeceqs}
& 1\,=(m-1)f''(1)\,\int_{h_0}^{\infty}dh\,\frac{P_{h_0}(h)}{[h+\Re x(\lambda_0)]^2} \\
& \lambda_0\,=\,(m-1)\Re x(\lambda_0)^2\,\int_{h_0}^{\infty}dh\,\frac{P_{h_0}(h)}{h\,[h+\Re x(\lambda_0)]^2}.\nonumber
\end{eqnarray}
We shall now consider the small $h_0$ limit of these last equations and the two cases $y>y_{mg}$ and $y=y_{mg}$. Let's begin with $y>y_{mg}$: the first integral in \ref{eq:gapSpeceqs} is dominated by the values of $h$ close to the cavity gap $h_0$; here $P_{h_0}(h_0)\sim h_0^{m-1}$, thus integrating in a small region $[h_0,c\,h_0]$ we get ($x_0\equiv x(\lambda_0)$)
\begin{eqnarray}
\label{eq:x_0scal}
& 1\sim \frac{(1-1/c)(-x_0)^{(m-1)}}{Z_{h_0}(h_0+x_0)} \nonumber \\
& x_0\,\sim\,-h_0+\frac{(1-1/c)}{Z_{h_0}}|x_0|^{m-1}
\end{eqnarray}
which ensures us that $|x_0|<h_0$. Then, rearranging the second of \ref{eq:gapSpeceqs}
\begin{equation*}
    \lambda_0\,=\,f''(1)\chi_{h_0}-x_0-f''(1)(m-1)\left\langle\frac{1}{h+x_0}\right\rangle_{h_0}
\end{equation*}
expanding $1/(h+x_0)$ in $x_0/h$ and simplifying:
\begin{equation*}
    \lambda_0\,=\,\Lambda|x_0|-f''(1)(m-1)x_0^2\left\langle\frac{1}{h^3}\right\rangle_{h_0}+O(x_0^3)
\end{equation*}
and plugging into this last equation eq. \eqref{eq:x_0scal}, it is found at leading order in $h_0$ 
\begin{equation}
    \lambda_0\,=\,\Lambda\,h_0+O(h_0^2)
\end{equation}
We consider now the case $y=y_{mg}$, i.e. $\Lambda=0$. Here one finds from the first of \eqref{eq:gapSpeceqs}
\begin{eqnarray}
\label{eq:gapSpecCrit}
    && \left\langle\frac{1}{h^2}\right\rangle_0\,=\,\left\langle\frac{1}{(h+x_0)^2}\right\rangle_{h_0}
\end{eqnarray}
which after a few manipulation yields
\begin{eqnarray}
\\
    && |x_0|\,=\begin{cases}
    \frac{h_0^{m-2}}{2\,(m-2)\,Z_0\,\langle 1/h^3 \rangle_0}+O(h_0^{m-1}),\,\,m>3 \\
    \\
    \frac{h_0}{2\,|\ln h_0|}+O(h_0^{2}),\,\,m=3
    \end{cases}
\end{eqnarray}
From the second of \eqref{eq:gapSpeceqs} then expanding $\Lambda_{h_0}$, setting $\Lambda=0$ and keeping terms up to order $x_0^2$, we find
\begin{equation}
    \lambda_0\,=\begin{cases}
    \left[\frac{f''(1)(m-1)}{4\, (m-2)^2\,Z_0^2\,\langle 1/h^3 \rangle_0}\right] h_0^{2(m-2)}+O(h_0^{2(m-1)}),\,\,\,m>3 \\
    \\
    \left[\frac{f''(1)}{2\, \,Z_0}\right] \frac{h_0^{2}}{|\ln h_0|}+O(h_0^{4}),\,\,\, m=3.
    \end{cases}
\end{equation}

We shall now consider the scaling of the spectral density and of the IPR close to $\lambda_0$. Equations \eqref{eq:lambdaIJ} are still valid if one replaces the ungapped $P_0(h)$ with the gapped one $P_{h_0}(h)$:
\begin{eqnarray*}
    && \lambda+\Lambda_{h_0}x\,=\,-x^2\,J_{h_0}-x|x|^2\,I_{h_0} \\
    && J_{h_0}(\lambda)\,=\,f''(1) (m-1) \left\langle\frac{1}{h |h+x|^2}\right\rangle_{h_0} \\
    && I_{h_0}(\lambda)\,=\,f''(1) (m-1) \left\langle\frac{1}{h^2 |h+x|^2}\right\rangle_{h_0}
\end{eqnarray*}
Differently from the gapless case, here the integrals $J_{h_0}$ and $I_{h_0}$ are always finite in the limit $\lambda\rightarrow\lambda_0$, for any $y_{mg}\leq y\leq y_{gs}$: at $\lambda=\lambda_0$, it follows directly from $h_0+x_0>0$. For $\lambda>\lambda_0$, one finds $h+x\simeq h+x_0-\frac{m-3}{m-2}(\lambda-\lambda_0)+\Im x$, since $d\Re x(\lambda_0)/d\,\lambda=-\frac{m-3}{m-2}$; so the integrals are well defined if and only $h_0+x_0+\Im x>\frac{m-3}{m-2}(\lambda-\lambda_0)$, so necessarily $\Im x\gg O(\lambda-\lambda_0)$. In fact, one finds that the spectral density has a square root behavior close to the spectral edge:

\begin{eqnarray}
    & \rho\simeq\sqrt{\frac{(1-C_{h_0})(\lambda-\lambda_0)}{J_{h_0}}} \\
    & C_{h_0}\,=\,|x_0|\left(\frac{m-3}{m-2}\right)[2\,J_0+|x_0|\,(dJ(x_0)/dx)] \\
    & J_{h_0}\,=\,f''(1)(m-1)\left\langle\frac{1}{h(h+x_0)^2}\right\rangle_{h_0}
\end{eqnarray}

As a consequence, the related lower edge eigenvectors of ultra-stable minima are found to be fully delocalised. Indeed, the IPR close to the spectral edge for $y>y_{mg}$ behaves as
\begin{eqnarray}
    && N\,IPR(\lambda)\propto \int_{h_0}^{\infty}\,\frac{dh\,P_{h_0}(h)}{|h+x|^4}\,=\,\int_{h_0}^{\infty}\,\frac{dh\,P_{h_0}(h)}{(h+x_0)^4} \nonumber \\ 
    && +O(\lambda-\lambda_0) \approx \frac{|x_0|^{m-1}}{3\,Z_0\,(h_0+x_0)^3}\sim h_0^{-2\,(m-1)}
\end{eqnarray}
At the critical point we find by similar manipulations
\begin{eqnarray}
    N\,IPR(\lambda_0)\sim \begin{cases}
    1/h_0,\,\,\,m=3 \\
    \ln h_0,\,\,\, m=4 \\
    const,\,\,\, m\geq 5.
    \end{cases}
\end{eqnarray}

\bibliography{HS_def}

\end{document}